\documentclass[aps, prd, amsmath, floats, floatfix, twocolumn, nofootinbib, superscriptaddress, showpacs]{revtex4-1}
\usepackage{graphicx,color}
\usepackage{hyperref}
\usepackage{latexsym}
\usepackage{amsmath,amsfonts,amssymb}
\usepackage{physics}     % convinient physics shortcuts

\begin{document}

% TITLE

\title{Stationary scalar clouds around a BTZ black hole}

% AUTHORS

\author{Hugo R. C. Ferreira}
\affiliation{Istituto Nazionale di Fisica Nucleare -- Sezione di Pavia, Via Bassi 6, 27100 Pavia, Italy}

\author{Carlos A. R. Herdeiro}
\affiliation{Departamento de F\'isica da Universidade de Aveiro and Center for Research and Development in Mathematics and Applications (CIDMA), Campus de Santiago, 3810-183 Aveiro, Portugal}

% DATE

\date{July 2017}

\begin{abstract}
We establish the existence of stationary clouds of massive test scalar fields around BTZ black holes. These clouds are zero-modes of the superradiant instability and are possible when Robin boundary conditions (RBCs) are considered at the AdS boundary. These boundary conditions are the most general ones that ensure the AdS space is an isolated system, and include, as a particular case, the commonly considered Dirichlet or Neumann-type boundary conditions (DBCs or NBCs). We obtain an explicit, closed form, resonance condition, relating the  RBCs that allow the existence of normalizable (and regular on and outside the horizon) clouds  to the system's parameters. Such RBCs \textit{never} include pure DBCs or NBCs. We illustrate the spatial distribution of these clouds, their energy and angular momentum density for some cases. Our results show that BTZ black holes with scalar hair can be constructed, as the non-linear realization of these clouds.
\vspace{13ex}
\end{abstract}

\maketitle

% INTRODUCTION

\section{Introduction}
The Kerr-Newman (KN) black holes (BHs) family~\cite{Kerr:1963ud,Newman:1965my} plays a central role in our understanding of BH physics. In electrovacuum, the ``uniqueness" theorems establish it as the only family of physically reasonable (single) BH solutions (see~\cite{Chrusciel:2012jk} for a review). Over the last few years, however, it has been shown that adding simple extra matter to the Einstein-Maxwell model, the KN family bifurcates to larger families of stationary, asymptotically flat, regular (on and outside the horizon) \textit{BHs with synchronized hair}~\cite{Herdeiro:2014goa,Herdeiro:2015gia,Kleihaus:2015iea,Herdeiro:2015tia,Chodosh:2015oma,Herdeiro:2016tmi,Delgado:2016jxq}, circumventing longstanding ``no-hair" theorems (see $e.g.$~\cite{Herdeiro:2015waa,Sotiriou:2015pka,Volkov:2016ehx}).

The existence of these ``hairy" BHs, bifurcating from the KN family, can be antecipated by considering the corresponding matter, in a test field approximation, on the Kerr(-Newman) background (see the discussion in~\cite{Herdeiro:2014ima}). As first observed by Hod~\cite{Hod:2012px}, and further developed in, $e.g.$~\cite{Hod:2013zza,Herdeiro:2014goa,Herdeiro:2015gia,Hod:2014baa,Benone:2014ssa,Herdeiro:2014pka,Hod:2014sha,Hod:2015goa,Hod:2015ynd,Hod:2016yxg,Hod:2016lgi}, under a certain \textit{resonance condition}, corresponding to a synchronization of the matter field's phase angular velocity with the horizon's angular velocity, \textit{real} frequency bound states of the corresponding matter field exist, dubbed \textit{stationary clouds around the BH}. The resonance condition corresponds precisely to the threshold of the superradiant instability of the corresponding ``bald" BH (see~\cite{Brito:2015oca} for a review), triggered by that matter field. Thus, these bound states are interpreted as \textit{superradiance zero-modes}, occurring in between decaying modes (into the BH) and superradiantly amplified modes (by the BH). It follows  that the hairy BHs may be regarded as the nonlinear realization of these stationary clouds, when their backreaction is taken into account and the fully nonlinear Einstein(-Maxwell)-matter system is solved. 

%The existence of these stationary clouds, moreover, guarantees that the family of  ``hairy BHs" is continuously connected to that of the ``bald (KN) BHs". 

One may ask if other well known BH solutions can equally be endowed with ``synchronized matter hair". A particularly interesting case, due to its simplicity, is the three dimensional BTZ black hole~\cite{Banados:1992wn,Banados:1992gq}. A major difference here, with respect to the aforementioned KN family, is that the BTZ BH is asymptotically anti-de-Sitter (AdS). This, however, is not an obstacle. In fact, the first example of a BH with synchronized (scalar) hair was found in a (five dimensional) asymptotically AdS spacetime~\cite{Dias:2011at}. Unlike its five dimensional counterpart, however, the geometry of the BTZ prevents the existence of superradiance for the simplest type of matter (a scalar field) and the simplest type of asymptotic boundary conditions [Dirichlet boundary conditions (DBCs)]~\cite{Ortiz:2011wd}, and the corresponding zero-mode is not present.

The purpose of this paper is to show that considering a more general type of boundary conditions at the AdS boundary --- Robin boundary conditions (RBCs), which are still totally reflective, thus preserving AdS as an isolated system ---, stationary clouds for a massive scalar field are possible. Our work follows the observation in~\cite{Iizuka:2015vsa} that superradiance exists when certain RBCs are imposed for a scalar field in BTZ. Here, we shall analyze in detail the occurrence of the stationary clouds, whose treatment can be performed entirely analytically, an attractive feature which for the Kerr case only occurs at extremality~\cite{Hod:2012px}.

\bigskip

The content of the present paper is as follows. In Section~\ref{nonextremal} we review the computation of the Klein-Gordon equation in the BTZ BH. In Section~\ref{bc} we discuss the most general boundary conditions that can be imposed on the matter field, compatible with regarding AdS as an isolated ``box". In Section~\ref{clouds} we obtain the requirement on the boundary conditions that yield stationary clouds and illustrate these clouds for specific sets of parameters. In Section~\ref{conclusions} we summarize our findings and present some final remarks. Throughout the paper we employ natural units in which $c = G_{\rm N} = \hbar = 1$ and a metric with signature $(-+++)$.

%%%%%%%%%%%%%%%%%%%%%%%%%%%
\section{Scalar field in the BTZ black hole}
\label{nonextremal}
%%%%%%%%%%%%%%%%%%%%%%%%%%%

The computation of the massive Klein-Gordon equation on the rotating BTZ BH is well known in the literature (see $e.g.$~\cite{Ghoroku:1994np,Ichinose:1994rg,Birmingham:2001hc}). The spacetime isometries allow full separation of variables (despite being a rotating BH) and the simplicity of the metric yields closed form solutions valid over the whole exterior spacetime, written in terms of hypergeometric functions. Let us review these solutions.

\subsection{BTZ black hole}

The metric of a BTZ BH in Schwarzschild-like coordinates is given by
\begin{align} \label{eQ:BTZmetric}
\dd s^2 = - N(r)^2 \dd t^2 + \frac{\dd r^2}{N(r)^2} + r^2 \left[\dd \phi + N^{\phi}(r) \dd t \right]^2 \, ,
\end{align}
where 
\begin{align}
N(r)^2 = - M + \frac{r^2}{\ell^2} + \frac{J^2}{4r^2} \, , \quad
N^{\phi}(r) = - \frac{J}{2r^2} \, ,
\end{align}
$M$ is the mass of the BH and $J$ is its angular momentum, whereas $\ell$ is the AdS radius. Observe that $\ell$ and $J$ have units of length whereas $M$ is dimensionless (which provides an interpretation for the absence of BH in three dimensional vacuum general relativity).

This BH solution has an event horizon at $r=r_+$ and an inner horizon at $r=r_-$, with $r_\pm$ being the roots of $N(r)^2$,
\begin{equation}
r_{\pm}^2 = \frac{\ell^2}{2} \left(M \pm \sqrt{M^2 - \frac{J^2}{\ell^2}}\right) \, .
\end{equation}
There is an ergoregion for $r_+ < r < r_{\rm erg} = \ell \sqrt{M}$, where $r_{\rm erg}$ is the radial coordinate of the ergo\textit{circle}. However, there is no speed of light surface, that is, a surface in the exterior region for which the Killing generator of the horizon, $\chi = \partial_t + \Omega_{\mathcal{H}} \partial_{\phi}$, is null, where
\begin{equation}
\Omega_{\mathcal{H}} = \frac{r_-}{\ell r_+} 
\end{equation}
is the angular velocity of the horizon. It will be convenient to rewrite the BH mass as a function of $\Omega_{\mathcal{H}}$,
\begin{equation} \label{eq:massBH}
M = \frac{r_+^2+r_-^2}{\ell^2} = \frac{r_+^2}{\ell^2} \left(1 + \ell^2 \Omega_{\mathcal{H}}^2 \right) \, .
\end{equation}
For completeness we also note that $J=2r_+r_-/\ell$.

The extremal BTZ BH is obtained by taking $|J|=M\ell$. Thus, the event and inner horizons coincide at $r_+ = r_- = \ell \sqrt{M/2}$ and the angular velocity of the horizon is, curiously, completely determined by the AdS radius, $\Omega_{\mathcal{H}} = 1/\ell$. In this case, the BH mass is related to $\Omega_{\mathcal{H}}$ by $M = 2 r_+ ^2\Omega_{\mathcal{H}}^2$.

\subsection{Klein-Gordon equation}

We consider a massive scalar field $\Phi$, with mass $\mu/\ell$, where $\mu$ is dimensionless, which satisfies the Klein-Gordon equation,
\begin{equation}
	\left(\nabla^2-\frac{\mu^2}{\ell^2} \right) \Phi = 0 \, ,
\end{equation}
and for which the mass satisfies the Breitenlohner-Freedman bound, $\mu^2 \geqslant -1$ \cite{Breitenlohner:1982jf}.

\subsubsection{Non-extremal case}

For the non-extremal BTZ BH, taking the ansatz 
\begin{equation}
\Phi(t,r,\phi) = e^{-i\omega t + ik\phi} \phi(r) \ ,
\label{ansatz}
\end{equation} 
introducing a new radial coordinate $z$ that compactifies the exterior region $r\in (r_+,\infty)$ into $z\in (0,1)$,  
\begin{equation}
z \equiv \frac{r^2-r_+^2}{r^2-r_-^2} \, , \qquad r_+\neq r_- \ ,
\label{zc}
\end{equation}
and letting $\phi(z)=z^\alpha(1-z)^\beta F(z)$, the radial equation transforms into the hypergeometric equation for $F(z)$. When\footnote{As we will see in the next section, when $\mu^2 = n^2-1, \ n \in \mathbb{N}$, no boundary conditions can be imposed at spatial infinity, so we will not consider this case further in this paper. The special case $\mu^2 = -1$ needs to studied separately and we will not pursue it in this paper.}
$\mu^2 \neq n^2-1, \ n \in \mathbb{N}_0$, two linearly independent solutions for $\phi(z)$ are
\begin{align}
	\phi^{\rm (D)}(z) &= z^{\alpha} (1-z)^{\beta} \notag \\
	&\quad \times F(a,b;a+b+1-c;1-z) \, , \label{sol1} \\
	\phi^{\rm (N)}(z) &= z^{\alpha} (1-z)^{1-\beta} \notag \\
	&\quad \times F(c-a,c-b;c-a-b+1;1-z) \, ,
	\label{sol2}
\end{align}
where
\begin{gather*}
	%\alpha \equiv  - i \ell \, \frac{\omega\ell r_+ - k r_-}{2 (r_+^2-r_-^2)} \, , \qquad
	\alpha \equiv  - i \, \frac{\ell^2 r_+ }{2 (r_+^2-r_-^2)}(\omega-k\Omega_{\mathcal{H}}) \, , \ \ \
	\beta \equiv \frac{1}{2}\left(1+\sqrt{1+\mu^2}\right) \, , \\
	a \equiv \beta - i\ell \, \frac{\omega\ell+k}{2(r_+ +r_-)} \, , \qquad
	b \equiv \beta - i\ell \, \frac{\omega\ell-k}{2(r_+ - r_-)} \, , \\
	c \equiv 1 + 2\alpha \, ,
\end{gather*}
and $F$ is the Gaussian hypergeometric function. The superscripts (D), (N) will become clear later. Observe that this general solution depends on six parameters: $r_-,r_+,\ell,\mu,k,\omega$.

In this paper, we shall be interested in obtaining stationary scalar modes, which is possible under a resonance condition for which the phase angular velocity of the mode $\omega/k$ equals the horizon angular velocity $\Omega_{\mathcal{H}}$,  
\begin{equation}
\omega = k \Omega_{\mathcal{H}} .
\label{sc}
\end{equation}
It follows that $\alpha=0$, $c=1$ and the solutions~\eqref{sol1}-\eqref{sol2} reduce to:
\begin{align}
	\phi^{\rm (D)}(z) &= (1-z)^{\beta} F(a,a^*;2\beta;1-z) \, ,\label{sol12} \\
	\phi^{\rm (N)}(z) &= (1-z)^{1-\beta} F(1-a,1-a^*;2-2\beta;1-z) \, , \label{sol22}
\end{align}
where $\beta$ is still as before but $a$ reduces to
\begin{gather*}
%	\beta = \frac{1}{2}\left(1+\sqrt{1+\mu^2}\right) \, , \\
	a = \beta - i\ell \, \frac{k}{2r_+} \, .
\end{gather*}
Interestingly, the general solution is now an explicit function of only \textit{four} parameters: $r_+,\ell,\mu,k$.
Moreover, note that both linearly independent solutions are now \emph{real}-valued solutions.

\subsubsection{Extremal case}

Let us now briefly discuss the extremal case. To solve the Klein-Gordon equation describing a massive scalar field $\Phi$, with $-1 \leqslant \mu^2 < 0$, we still take the ansatz~\eqref{ansatz} but replace the compactified radial coordinate~\eqref{zc} by 
\begin{equation}
z \equiv \frac{r_+^2}{r^2-r_+^2} \, .
\end{equation}
This $z$ coordinate is \textit{non-compact} and maps the exterior region $r\in (r_+,\infty)$ into $z\in (0,+\infty)$, with the AdS boundary at $z\rightarrow 0$.  

Two linearly independent mode solutions are
\begin{align}
\Phi^{\rm (D)}(z) &= z^{\beta} e^{i \alpha_- z} M(a,b,-2i\alpha_-z) \, , \\
\Phi^{\rm (N)}(z) &= z^{1-\beta} e^{i \alpha_- z} M(a-b+1,2-b,-2i\alpha_-z) \, ,
\end{align}
where $M(a,b,z)$ is the Kummer's confluent hypergeometric function, with $\beta$ as before and
\begin{gather*}
\alpha_{\pm} \equiv \ell \, \frac{\omega\ell \pm k}{2r_+} \, , 
\qquad
%\beta = \frac{1}{2}\left(1+\sqrt{1+\mu^2}\right) \, , \\
a \equiv \beta - \frac{i}{2} \alpha_+ \, , \qquad
b \equiv 2\beta \, .
\end{gather*}
The first one is the solution that satisfies the DBC at $z=0$ (i.e.~it is the principal solution -- as defined below -- at $z=0$), whereas the second one satisfies a NBC.

Imposing the resonance condition~\eqref{sc}, which now simplifies to $\omega = k \Omega_{\mathcal{H}} = k/\ell$, the solutions simplify considerably and become
\begin{align} \label{eq:extremalsolutions}
\Phi^{\rm (D)}(z) = z^{\beta}  \, , \qquad
\Phi^{\rm (N)}(z) = z^{1-\beta} \, .
\end{align}
In the extremal case, synchronised solutions depend on a \textit{single} parameter, $\mu$.

% As before, to impose the regularity condition on the horizon, we introduce another set of two linearly independent mode solutions:
%
%\begin{align}
%\Phi^{\rm in}(z) &= z^{\beta} e^{i \alpha_- z} U(a,b,-2i\alpha_-z) \, ,\label{extremalreg} \\
%\Phi^{\rm out}(z) &= z^{\beta} e^{-i \alpha_- z} U(b-a,b,2i\alpha_-z) \, ,
%\end{align}
%
%where $U(a,b,z)$ is the Tricomi's confluent hypergeometric function. 

% As in the non-extremal case, requiring regularity at the horizon when $\omega = - k \Omega_{\mathcal{H}} = -k/\ell$, amounts to considering solution~\eqref{extremalreg} \ch{[Hugo, please check or correct this.]} and a similar analysis as before leads to the resonance condition  $\zeta = \zeta_{\rm ex}$. \ch{Am I understanding this correctly? Could you expand a little here?}

%%%%%%%%%%%%%%%%%%%%%%%
\section{Robin boundary conditions}
\label{bc}
%%%%%%%%%%%%%%%%%%%%%%%

The AdS timelike (conformal) boundary yields the possibility of placing material sources (or absorbers) on the boundary. Thus, different boundary conditions with different physical implications are possible. Here, we wish to regard the AdS spacetime, containing the matter field and BH, as an isolated system. In this section, we show  this requires that one considers generic Robin boundary conditions (RBCs). We remark that the implications of non-DBCs on the field propagation in asymptotically AdS spacetimes have been considered, $e.g.$, in~\cite{Dias:2013sdc,Cardoso:2013pza,Wang:2015goa,Wang:2015fgp,Wang:2016dek,Wang:2016zci}.

Consider a massive scalar field $\Phi$ propagating on the BTZ BH. Therein, we construct two linearly independent mode solutions of the Klein-Gordon equation, $\Phi^{\rm (D)}(t,r,\phi)$ and $\Phi^{\rm (N)}(t,r,\phi)$. $\Phi^{\rm (D)}$ is chosen to be the \textit{principal solution} at $r \to \infty$, that is, the unique solution (up to scalar multiples) such that $\lim_{r \to \infty} \Phi^{\rm (D)}(t,r,\phi)/\Psi(t,r,\phi) = 0$ for every solution $\Psi$ that is not a scalar multiple of $\Phi^{\rm (D)}$. 

The asymptotic behavior of the pair of solutions \eqref{sol1}-\eqref{sol2} as $z \to 1$ ($r \to \infty$) is as follows
\begin{align}
\phi^{\rm (D)}(z) &\sim (1-z)^{\frac{1}{2}\left(1+\sqrt{1+\mu^2}\right)} \sim r^{-1-\sqrt{1+\mu^2}} \, , \\
\phi^{\rm (N)}(z) &\sim (1-z)^{\frac{1}{2}\left(1-\sqrt{1+\mu^2}\right)} \sim r^{-1+\sqrt{1+\mu^2}} \, .
\end{align}
It is easy to see that $\phi^{\rm (D)}$ is the radial part of the desired principal solution $\Phi^{\rm (D)}$. This is the \textit{Dirichlet solution}. The other solution, $\Phi^{\rm (N)}$, is a nonprincipal solution and it is not unique, as any linear combination of this solution and the principal solution is another nonprincipal solution. We shall call it the \textit{Neumann solution}. A general solution may, in principle, be written as $\Phi = C^{\rm (D)} \Phi^{\rm (D)} + C^{\rm (N)} \Phi^{\rm (N)}$, where $C^{\rm (D)}$ and $C^{\rm (N)}$ are two complex constants.

For convenience, we introduce another set of linearly independent solutions,
\begin{equation}
\Phi_+ = \Phi^{\rm (D)} - i \Phi^{\rm (N)} \, , \quad
\Phi_- = \Phi^{\rm (D)} + i \Phi^{\rm (N)} \, ,
\end{equation}
such that a general solution is written as $\Phi = C_+ \Phi_+ + C_- \Phi_-$, where $C_+$ and $C_-$ are two complex constants.

The flux of energy at $r \to \infty$ is given by
\begin{equation}
\mathcal{F} = \lim_{r \to \infty} \int_{\Sigma_r} \dd \phi \sqrt{-g} \, g^{rr} T_{rt} \, ,
\end{equation}
where $\Sigma_r$ is a hypersurface of constant $r$ and $T_{\mu\nu}$ is the energy-momentum tensor of the scalar field. 
This can be computed and the result is
\begin{equation}
\mathcal{F} \propto \left(|C_+|^2 - |C_-|^2 \right) \, .
\end{equation}
Following the physical principle that the system is isolated ($i.e.$ there are no sources or sinks at the boundary), we require vanishing flux at infinity, which implies $|C_+| = |C_-|$. As a consequence, if we write $C_{\pm} = \rho \, e^{i \theta_{\pm}}$, we have, for $C^{\rm (D)} \neq 0$,
\begin{align}
	\frac{C^{\rm (N)}}{C^{\rm (D)}} &= -i \, \frac{C_+ - C_-}{C_+ + C_-} = \tan\left(\frac{\theta_+-\theta_-}{2}\right) \notag \\
	&\equiv \tan(\zeta) \in \mathbb{R} \, , \quad \zeta \in [0,\pi) \setminus \{\tfrac{\pi}{2}\} \, .
\end{align}
Hence, the scalar field has to satisfy RBCs in order for the flux to be zero at infinity. It can then be written as
\begin{equation} \label{eq:RBC}
\Phi = \cos(\zeta) \Phi^{\rm (D)} + \sin(\zeta) \Phi^{\rm (N)} \, , \quad \zeta \in [0,\pi) \, .
\end{equation}
This is the form we shall use in the following sections. Observe that the (most standard) Dirichlet boundary conditions (DBCs) corresponds to $\zeta=0$.

To close this section, we obtain the range of $\mu^2$ for which it is possible to apply RBCs. In short, these boundary conditions can be applied for the values of $\mu^2$ for which both linearly independent solutions are square-integrable near infinity \cite{Dappiaggi:2016fwc}. Note that $\phi^{\rm (D)}$ is square-integrable near infinity for all $\mu^2 > - 1$, that is,
\begin{equation}
\int^{\infty} \dd r \sqrt{-g} \, g^{tt} \big|\phi^{\rm (D)}(r)\big|^2 < \infty \, .
\end{equation}
As for the Neumann solution $\phi^{\rm (N)}$, it is square-integrable near infinity for $-1 < \mu^2 < 0$. If $\mu^2 \geqslant 0$, then only the solution $\phi^{\rm (D)}$ is square-integrable near infinity and no boundary conditions need to be imposed. 

In conclusion, RBCs may be applied for scalar fields with mass parameter such that $-1 < \mu^2 < 0$ and no boundary conditions are applied if $\mu^2 \geqslant 0$. Observe, in particular, that in the massless case $\mu^2=0$ no RBCs may be imposed for normalizable modes.

%%%%%%%%%%%%%%%%%%%%%%
\section{Stationary clouds}
\label{clouds}
%%%%%%%%%%%%%%%%%%%%%%

Physical (scattering, quasi-bound or quasinormal) modes satisfy ingoing boundary conditions at the horizon. For the problem of \textit{bound states} that we consider here, however,  the correct boundary condition at the horizon is decided based on \textit{regularity}. To see this, it is convenient to consider another set [different from~\eqref{sol12}-\eqref{sol22}] of linearly independent solutions for the non-extremal BH,
\begin{align}
\phi(z) &= A \, (1-z)^{\beta} F(a,a^*;1;z) 
+ B \, (1-z)^{\beta} \notag \\
&\quad \times \left\{ F(a,a^*;1;z) \log(z) + \sum_{j=1}^{\infty} z^j f(j) \right\} \, .
\end{align}
where
\begin{align*}
f(j) &= \frac{(a)_j(a^*_j)}{(j!)^2} \left[\psi(a+j) - \psi(a) + \psi(a^*+j) - \psi(a^*) \right. \notag \\
&\hspace{12ex} \left. - 2\psi(j+1) + 2 \psi(1) \right] \, ,
\end{align*}
and $(a)_j = \Gamma(a+j)/\Gamma(a)$ and $\psi$ is the digamma function.

The first term has a polynomial expansion near $z=0$, whereas the second term is logarithmically divergent as $z \to 0$. Hence, regularity at the horizon requires $B=0$. As pointed out above, the behavior of the scalar field near the horizon is \textit{not} a wave-like behavior. The synchronization condition~\eqref{sc} changes the near-horizon scalar equation, changing the wave-like solution by a polynomial expansion. This ensures there is no flux towards (or from) the horizon, hence explaining why one may find bound states (with a real frequency) rather than merely quasi-bound states (with a complex frequency).

In the extremal case, there is no linear combination of the solutions \eqref{eq:extremalsolutions} which is regular at the horizon, $z \to \infty$. Therefore, there are no stationary scalar cloud configurations around extremal BTZ BHs: there is a discontinuous behaviour of the stationary clouds, at the extremal BTZ limit. A discontinuity that bears some resemblance has been recently discussed for zero damping quasinormal modes for the extremal Kerr BH~\cite{Richartz:2017qep}.

Returning to the non-extremal case, in order to relate this solution to the previously obtained ones \eqref{sol12}-\eqref{sol22}, we perform the transformation $z \to 1-z$ of the hypergeometric function \cite{NIST} and obtain
\begin{align*}
\phi(z) = A \left[ \frac{\Gamma(1-2\beta) \, \phi^{\rm (D)}(z)}{\Gamma(1-a)\Gamma(1-a^*)}  + \frac{\Gamma(2\beta-1) \, \phi^{\rm (N)}(z)}{\Gamma(a)\Gamma(a^*)}  \right] \, .
\end{align*}
Comparing with \eqref{eq:RBC}, one obtains
\begin{align} 
\tan(\zeta) &= \Xi(\mu^2,k,r_+,\ell) \, ,
\label{rc}
\end{align}
where
\begin{multline} 
\Xi(\mu^2,k,r_+,\ell) = \frac{\Gamma(2\beta-1)\Gamma(1-a)\Gamma(1-a^*)}{\Gamma(1-2\beta)\Gamma(a)\Gamma(a^*)} \\
= \frac{\Gamma\left(\sqrt{1+\mu^2}\right)\left|\Gamma\left(\frac{1}{2}-\frac{1}{2}\sqrt{1+\mu^2}+i\frac{k\ell}{2r_+}\right)\right|^2}{\Gamma\left({-\sqrt{1+\mu^2}}\right)\left|\Gamma\left(\frac{1}{2}+\frac{1}{2}\sqrt{1+\mu^2}+i\frac{k\ell}{2r_+}\right)\right|^2} \, .
\label{eq:RBCQNM}
\end{multline}
Eq.~\eqref{rc} is the \textit{resonance condition for scalar stationary clouds around non-extremal BTZ BHs}. Fixing the scalar field mass, the background parameters and the cloud quantum number $k$ fixes the right hand side of Eq.~\eqref{rc} and hence the value of $\zeta$ that defines the RBC that can yield that cloud. As a check on eq.~\eqref{eq:RBCQNM}, it reproduces the particular example considered in \cite{Iizuka:2015vsa}: for $\mu^2 = -8/9$, $k=1$, $\ell=1$, $r_+=5$ and $r_-=3$ we obtain $\cot(\zeta)=-0.414$, which coincides with the value presented therein.

An analysis of the resonance condition shows that, for $-1 < \mu^2 < 0$, the allowed values of $\zeta$ fall in the domain $[\zeta_*,\pi)$, where
\begin{equation*}
\zeta_* = \arctan(\frac{\Gamma\left(\sqrt{1+\mu^2}\right)\Gamma\left(\frac{1}{2}-\frac{1}{2}\sqrt{1+\mu^2}\right)}{\Gamma\left({-\sqrt{1+\mu^2}}\right)\Gamma\left(\frac{1}{2}+\frac{1}{2}\sqrt{1+\mu^2}\right)}) 
\end{equation*}
is such that $\zeta_* \in (\frac{\pi}{2},\pi)$. In other words, there are no cloud configurations for RBCs with $\zeta \in [0,\zeta_*)$, which in particular includes pure DBCs and NBCs, in agreement with previous results~\cite{Ortiz:2011wd}. 

Another perspective on the resonance condition is that fixing the scalar field parameters $\mu^2$, $\zeta$ and $k$, and for a given $r_+$ or $\ell$, stationary clouds only exist for a discrete set of values of $J$. As an illustration, in Fig.~\ref{fig:M-Omega-fixedrp}
% and \ref{fig:M-Omega-fixedl} 
we display some examples of \textit{existence lines} for the stationary clouds, in an $M$ versus $\Omega_{\mathcal{H}}$ diagram. In particular, comparing the left panel with the same type of plot for the Kerr case (see Fig. 1 in~\cite{Herdeiro:2014goa}), one verifies significant differences: in the Kerr case $M=1/(2\Omega_\mathcal{H})$ for extremal BHs and non-extremal BHs exist \textit{below} this extremal line; for the BTZ case $M=2\Omega_\mathcal{H}$ for extremal BHs and non-extremal BHs exist \textit{above} this extremal line.

In Fig.~\ref{fig:phi-z} we illustrate the radial profile of a selection of clouds. It is worth noticing that, as we vary the value of $\zeta$ (and correspondingly $\ell$) for fixed $\mu^2$ and $k$, the radial profile of the stationary clouds can change qualitatively. 
% In particular, it may or may not have nodes.
%
In Fig.~\ref{fig:phi-z} we also show the energy density and angular momentum density of the same cases for which the radial profile is plotted, using the appropriate components of the energy-momentum tensor associated to the scalar field $\Phi$, which is given by
\begin{equation}
T_{\mu\nu} = 2 \partial_{(\mu} \Phi^* \partial_{\mu)} \Phi - g_{\mu\nu} \left(\partial_{\lambda} \Phi^* \partial^{\lambda} \Phi + \mu^2 \ell^2 \Phi^* \Phi \right) \, .
\end{equation}
From these plots one can see that both the radial profiles as well as the energy and angular momentum distributions are everywhere regular and smooth.

\begin{widetext}

\begin{figure}[t!]
	\centering
	\includegraphics[width=0.44\linewidth]{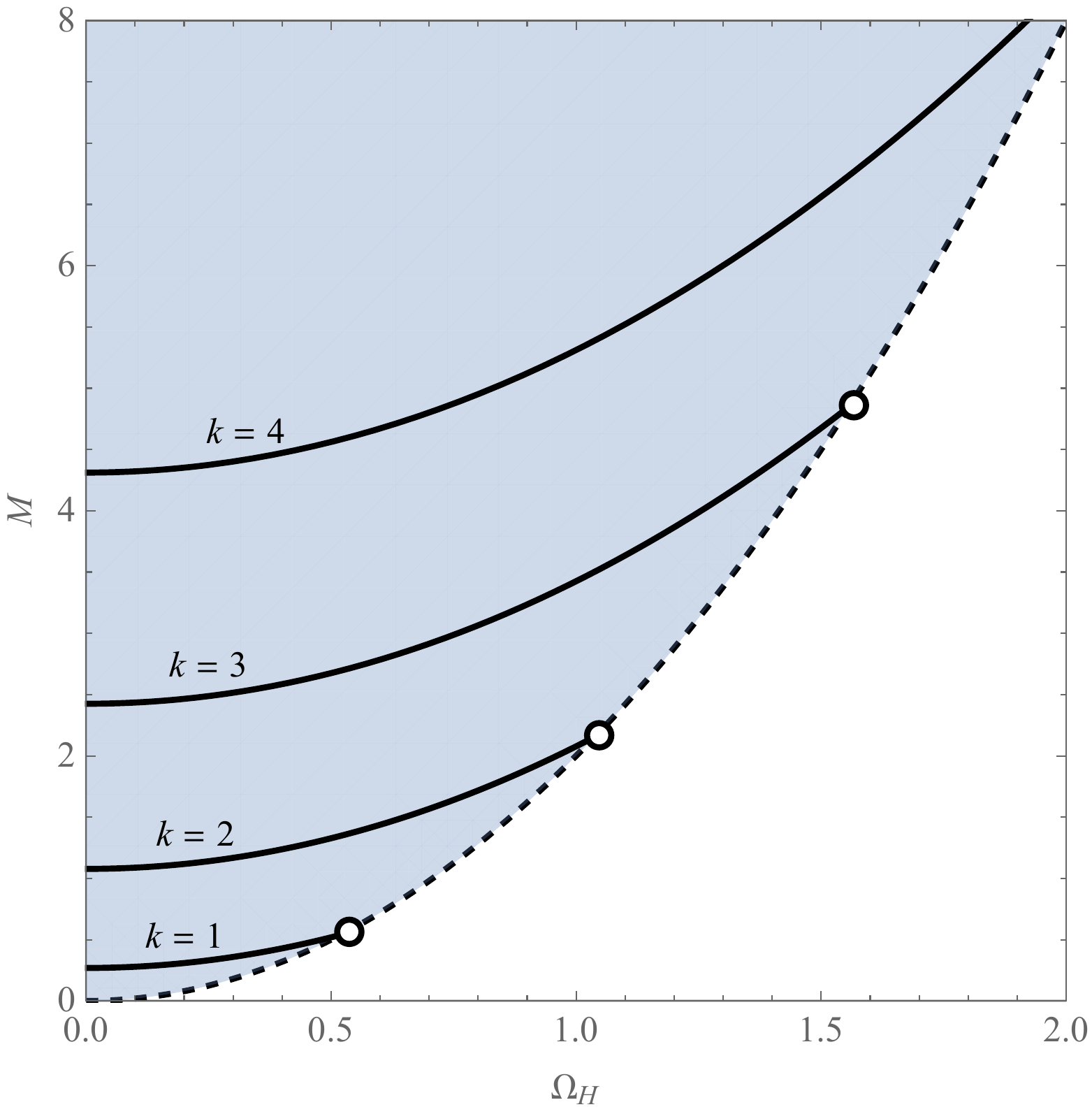} \hspace{8ex}
	\includegraphics[width=0.44\linewidth]{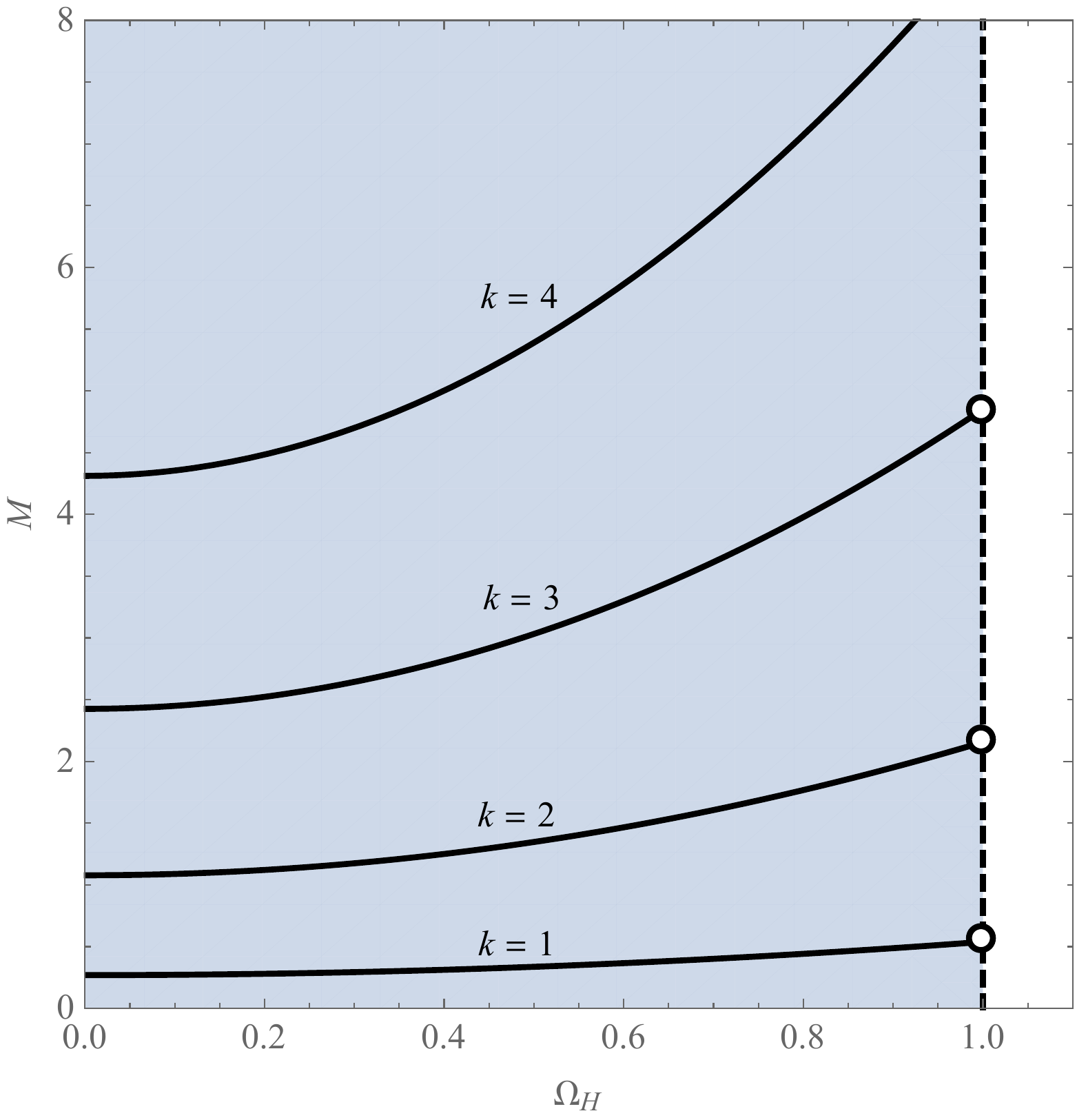}
	\caption{\label{fig:M-Omega-fixedrp}
	Stationary scalar clouds with $\mu^2 = -1/2$, $\zeta = 9\pi/10$ and $k=1,2,3,4$ (solid lines) on a  $M$ versus $\Omega_{\mathcal{H}}$ plot, for BTZ BHs with $r_+=1$ (left panel) or  $\ell=1$ (right panel). The dashed black curve corresponds to extremal BTZs, for which $\Omega_{\mathcal{H}} = 1/\ell$ and $M = 2 \Omega_{\mathcal{H}}$ (left panel) or $\Omega_{\mathcal{H}} = 1$ (right panel); non-extremal BTZ BHs exist in the shaded region. Each different line correspond to a different value of $\ell$ (left panel) or $r_+$ (right panel).}
\end{figure}

\begin{figure}[t!]
	\centering
	\includegraphics[width=0.45\linewidth]{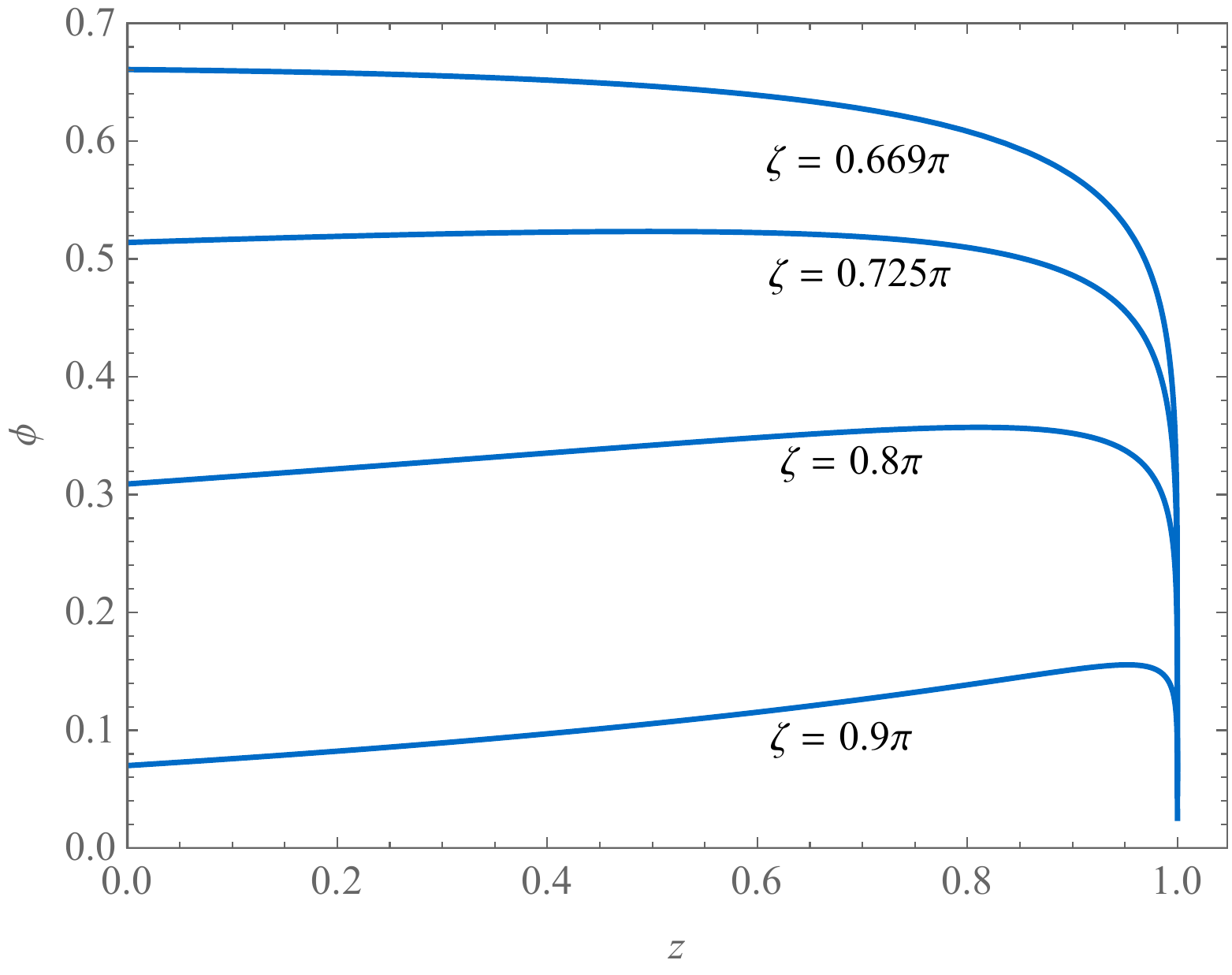} \hspace{5ex}
	\includegraphics[width=0.45\linewidth]{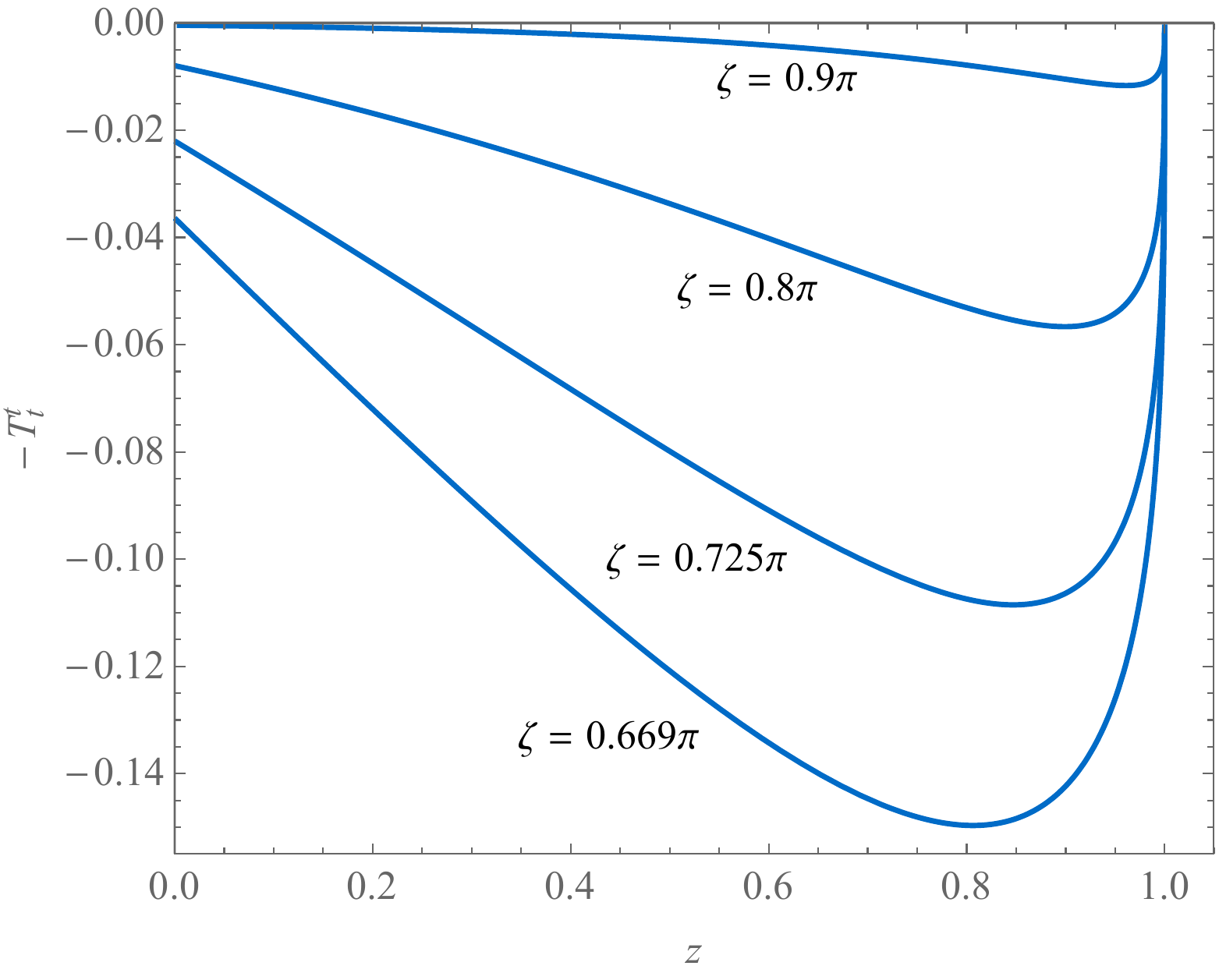} \\ \vspace*{4ex}
	\includegraphics[width=0.45\linewidth]{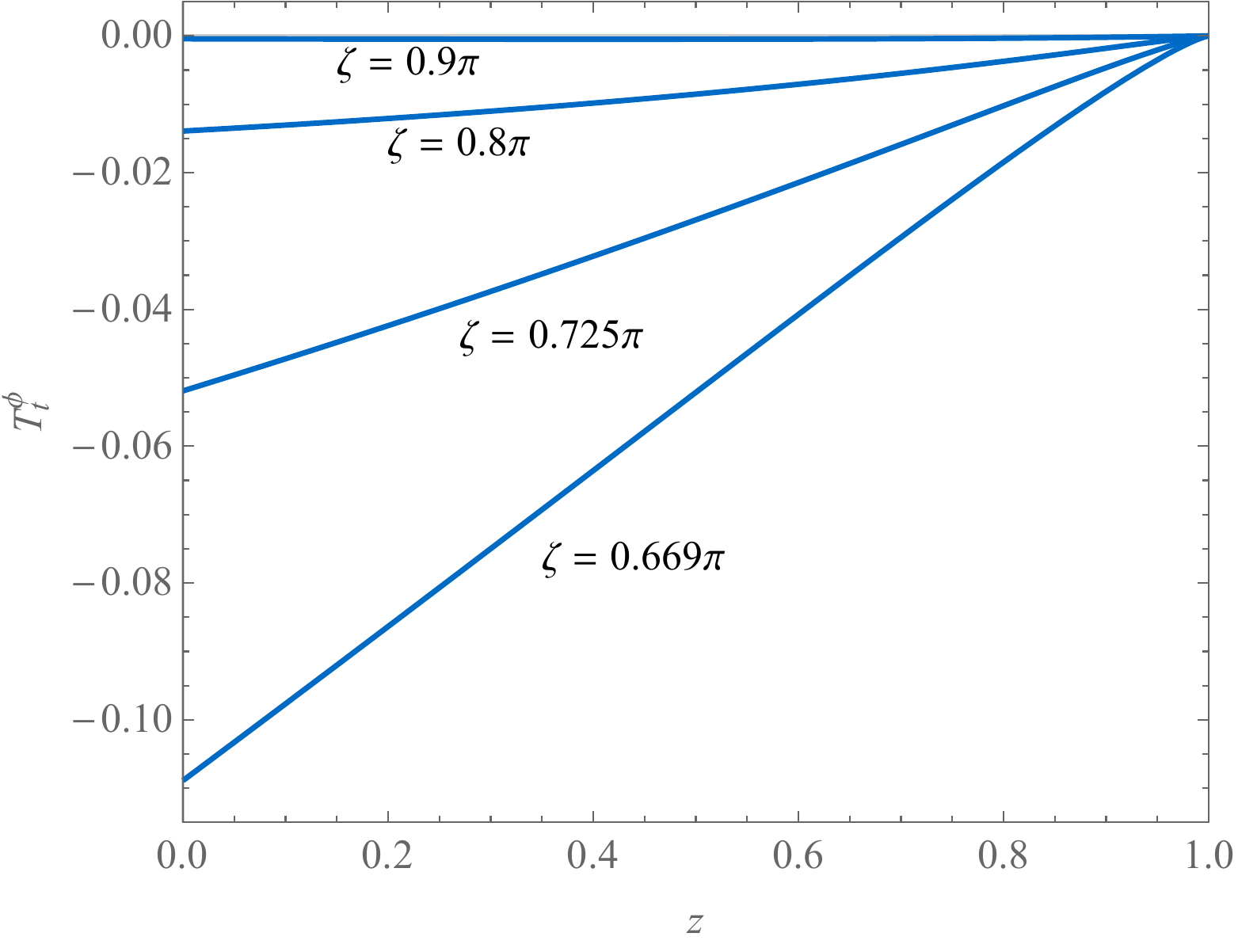}
	\caption{\label{fig:phi-z}
	Stationary scalar clouds with $\mu^2 = -1/2$ and $k=1$ on a $\phi$ versus $z$ plot (top left panel), a $-{T_t}^t$ versus $z$ plot (top right panel) and a ${T_t}^{\phi}$ versus $z$ plot (bottom panel), for BTZ BHs with $r_+=2$ and $r_-=1$, for different RBCs at infinity (and correspondingly different values of $\ell$). For a scalar field with this mass, the minimum value of $\zeta$ for which there are stationary clouds is $\zeta_* \approx 0.66876\pi$.}
\end{figure}

%
%\begin{figure}[t!]
%	\centering
%	\includegraphics[width=0.3\linewidth]{Figs/M-Omega-fixedl}
%	\caption{\label{fig:M-Omega-fixedl}$M$ versus $\Omega_{\mathcal{H}}$ for BTZ BHs with $\ell=1$. The solid black curve corresponds to extremal BTZs, for which $\Omega_{\mathcal{H}} = 1$; non-extremal BTZ BHs exist in the shaded region. Scalar clouds with $\mu^2 = -1/2$, $\zeta = 9\pi/10$ and $k=1,2,3,4$ exist along the dashed lines. Each different line correspond to a different value of $r_+$.}
%\end{figure}

\end{widetext}

% CONCLUSIONS

%%%%%%%%%%%%%%%%%%%%%%%%%%%
\section{Conclusions}
\label{conclusions}
%%%%%%%%%%%%%%%%%%%%%%%%%%%

The BTZ BH~\cite{Banados:1992wn,Banados:1992gq} stands out as a simple, geometrically elegant, BH solution of three dimensional general relativity (with a negative cosmological constant). In this paper we have shown that using appropriate RBCs, BTZ BHs can support \textit{stationary} scalar clouds of a massive scalar field. The stationarity of the clouds means that their frequency is real, and actually, synchronized with the BH horizon angular velocity, through relation~\eqref{sc}. For a complex scalar field, the corresponding energy momentum tensor will be invariant under the Killing vector fields $\partial/\partial t$ and $\partial/\partial \phi$. Hence, the backreaction of the clouds can (and should~\cite{Herdeiro:2014ima}) yield a family of stationary and axisymmetric BTZ BHs with synchronized scalar hair. We hope to report on the construction of these solutions in the near future,\footnote{Such solutions will have a solitonic limit. Examples of gravitating solitons (boson stars) in three dimensional AdS spacetime have been constructed in~\cite{Astefanesei:2003rw,Stotyn:2013spa}.} but we remark that these are different from the example discussed in~\cite{Iizuka:2015vsa}, wherein the geometry is invariant under a \textit{single} Killing vector field. A quite different example of a ``hairy" BTZ BH has been reported in~\cite{Harms:2017yko}, using non-linear sigma models.

The RBCs are fundamental for the existence of the stationary clouds reported here. If the more standard DBCs are imposed, without imposing the synchronization condition~\eqref{sc}, it can be observed that, generically, only quasi-bound states exist (with a complex frequency). However, taking the extremal BTZ limit, for one branch of quasi-bound states, the imaginary part vanishes and the real part synchronzes, $i.e.$ reduces to~\eqref{sc}. A very analogous type of behaviour has been observed for charged BHs in~\cite{Degollado:2013eqa,Sampaio:2014swa} (replacing the horizon angular velocity by the horizon electrostatic potential, and $k$ by the field's charge), where they have been dubbed \textit{marginal clouds around BHs}\footnote{We would like to thank M. Wang for this observation.} --- see also~\cite{Hod:2016jqt,Hod:2017gvn}. 

Finally, we would like to mention two possible continuations of this work.  Firstly, the results in this paper suggest detailed studies of superradiant instabilities of the BTZ BH, triggered by scalar fields with RBCs, could be quite interesting. We remark that superradiance was argued to occur in the BTZ background in~\cite{Carlip:1995qv}, motivated by considerations of quantum field theory on this background; but as mentioned above, under DBCs superradiance does not occur~\cite{Ortiz:2011wd}. The observation in~\cite{Iizuka:2015vsa}  together with our work invite us to revisit this problem, considering the more general class of RBCs (see also~\cite{Winstanley:2001nx} for related remarks on the relevance of boundary conditions for the occurrence (or not) of superradiance on four dimensional Kerr-AdS ).   Secondly, since the BTZ BH arises as identifications of three dimensional AdS spacetime, it would also be interesting to understand if and how the stationary clouds we have presented here are related to AdS normal modes.

\bigskip

% ACKNOWLEDGEMENTS

\begin{acknowledgments} %THANK INFN PAVIA CARLOS!
We would like to thank C.~Dappiaggi, E.~Radu and M.~Wang for discussions and E.~Winstanley for comments on a draft of this paper.  C.H. is grateful to the INFN -- Sezione di Pavia for the kind hospitality during the realization of part of this work. C.~H. acknowledges funding from the FCT-IF programme. This work was partially supported by the H2020-MSCA-RISE-2015 Grant No. StronGrHEP-690904, and by the CIDMA project
UID/MAT/04106/2013. 
The work of H.~F. was supported by the INFN postdoctoral fellowship ``Geometrical Methods in Quantum Field Theories and Applications''.
\end{acknowledgments}

% BIBLIOGRAPHY

\end{document}